\documentclass[english,aps,prb,reprint,showpacs]{revtex4-1}
\usepackage[T1]{fontenc}
\usepackage[latin9]{inputenc}
\usepackage{color}
\usepackage{babel}
\usepackage{units}
\usepackage{amstext}
\usepackage{amssymb}
\usepackage{graphicx}
\usepackage[unicode=true,
 bookmarks=true,bookmarksnumbered=false,bookmarksopen=false,
 breaklinks=false,pdfborder={0 0 1},backref=false,colorlinks=true]
 {hyperref}
\hypersetup{
 pdfauthor={Tomasz M. Kott}}
\usepackage{breakurl}

\makeatletter
 
 \@ifundefined{textcolor}{}
 {%
   \definecolor{BLACK}{gray}{0}
   \definecolor{WHITE}{gray}{1}
   \definecolor{RED}{rgb}{1,0,0}
   \definecolor{GREEN}{rgb}{0,1,0}
   \definecolor{BLUE}{rgb}{0,0,1}
   \definecolor{CYAN}{cmyk}{1,0,0,0}
   \definecolor{MAGENTA}{cmyk}{0,1,0,0}
   \definecolor{YELLOW}{cmyk}{0,0,1,0}
 }

\makeatother

\begin{document}

\title{Valley Degenerate 2D Electrons in the Lowest Landau Level}

\author{Tomasz M. Kott}

\email{tkott@mailaps.org}

\author{Binhui Hu}

\author{S. H. Brown}

\author{B. E. Kane}

\affiliation{Laboratory for Physical Sciences \& Joint Quantum Institute, University
of Maryland, College Park, MD 20740}

\date{\today}
\begin{abstract}
We report low temperature magnetotransport measurements on a high
mobility ($\mu=\unit[325\,000]{cm^{2}/V\, sec}$) 2D electron system
on a H-terminated Si(111) surface. We observe the integral quantum
Hall effect at all filling factors $\nu\leq6$ and find that $\nu=2$
develops in an unusually narrow temperature range. An extended, exclusively
even numerator, fractional quantum Hall hierarchy occurs surrounding
$\nu=3/2$, consistent with two-fold valley-degenerate composite fermions
(CFs). We determine activation energies and estimate the CF mass.
\end{abstract}


\maketitle
Multicomponent two dimensional (2D) systems 
in a single Landau level have generated interest
due to the possibilities for novel correlated ground states in the
integer and fractional quantum Hall (FQH) regimes when the energies
of the component states become degenerate. Early experiments focused
on measurements of engineered GaAs materials where the spin splitting
could be reduced to zero \cite{leadley1997fractional,kang1997evidence,shukla2000largeskyrmions}.
Subsequent development of AlAs quantum wells with a tunable valley
degeneracy allowed the study of a spin-like degeneracy in the same
limit \cite{shayegan2006twodimensional}. Recently, there has been
great interest in the sublattice (valley) degeneracy in graphene;
experiments show that the valley symmetry affects the FQH hierarchy
\cite{dean2011multicomponent,feldman2012unconventional2} and that
valley ferromagnetism occurs when one of two degenerate valleys is
occupied \cite{young2012spinand}. 

Measurements on silicon, the first multi-valley system to be considered
theoretically \cite{rasolt1985newgapless,rasolt1986dissipation},
had been hampered by high disorder. Lately, however, Si(100)/SiGe
heterostructures have shown mobilities up to $\unit[10^{6}]{cm^{2}/Vs}$
\cite{lu2012fractional}. Nonetheless, Si(100) is known to have an
intrinsic valley splitting due to the confinement potential
\cite{boykin2004valleysplitting,tsui1979observation,takashina2006valleypolarization,goswami2007controllable}.
The case of 2D electrons on (111) oriented silicon surfaces, which
have three pairs of opposite momentum ($\vec{k}$) valleys, is especially
interesting \cite{hwang2012valley}. As opposed to either AlAs or Si(100), 
the degeneracy of valley pairs with
$\pm\vec{k}$ symmetry in Si(111) cannot be broken within the effective
mass approximation or by a confinement potential \cite{rasolt1986dissipation},
similar to the case of valleys in graphene. Additionally, both AlAs
and Si(111) exhibit anisotropic mass tensors; in AlAs, this anisotropy
arguably transfers to composite fermions (CFs) \cite{gokmen2010transference}.
Novel broken symmetry states have been predicted at integer filling
factor, $\nu$, when the Fermi energy, $E_{F}$, lies between two
valleys with different mass tensors \cite{abanin2010nematic}. 

In this report we present transport data on a very high mobility electron
system ($\mu=\unit[325\,000]{cm^{2}/Vs}$ at a temperature $T=\unit[90]{mK}$
and density $n_{s}=\unit[4.15\times10^{11}]{cm^{-2}}$) on a hydrogen-terminated
Si(111) surface. We observe an extended FQH hierarchy around $\nu=3/2$.
We argue that the FQH hierarchy is consistent with the SU(2) symmetry
of a two-fold valley-degenerate ground state and present the 
first measurement of the CF mass in a multicomponent system, as
well as the first in an anisotropic system. 
We also present preliminary evidence for many-body interactions
affecting integer activation energies: the development of $\nu=2$ 
occurs in an unusually narrow temperature range, which may signal 
a transition to broken symmetry valley states. 

\begin{figure}
\centering{}\includegraphics[width=3.375in]{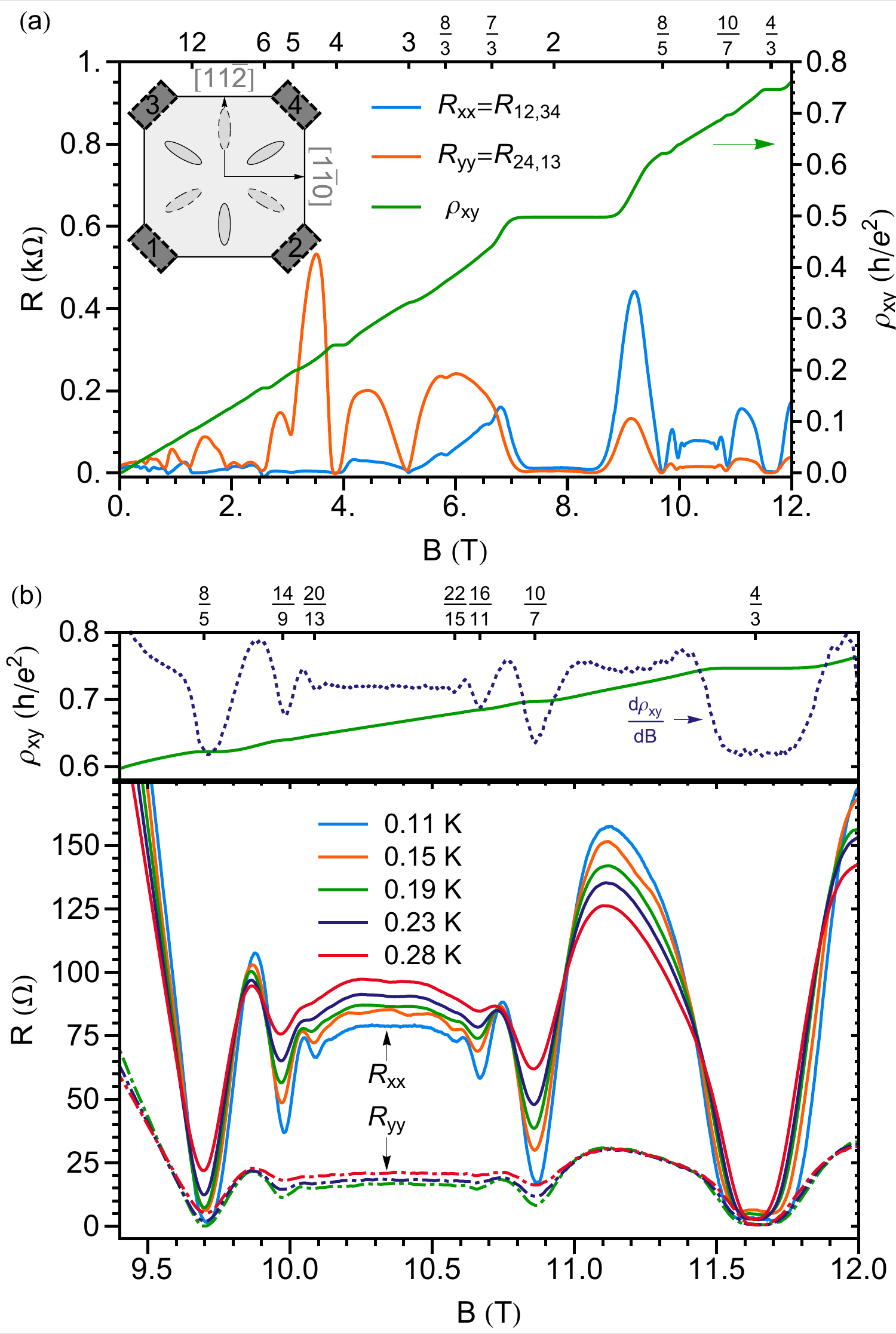}
\caption{(Color) \emph{(a)} $R_{xx}$, $R_{yy}$ (left), and $\rho_{xy}$ (right) vs.~$B$
at $T=\unit[90]{mK}$ and $n_{s}=\unit[3.75\times10^{11}]{cm^{-2}}$
with filling factors listed on the top axis; the fractional states
at $\nu=\frac{8}{5},\,\frac{10}{7},$ and $\frac{4}{3}$ have plateaus
in $\rho_{xy}$. \emph{(b)} $R_{xx}$ and $R_{yy}$ vs.~$B$ for 
$n_{s}=\unit[3.75\times10^{11}]{cm^{-2}}$.
Seven FQH states are visible at $T=\unit[110]{mK}$; the filling factors
are labeled on the top axis. \emph{Top panel}: $\rho_{xy}$ versus
$B$ for $T=\unit[110]{mK}$. The derivative is also shown to accentuate
the fine structure, matching the $R_{xx}$ minima. 
\label{fig:bSweep}}
\end{figure}
For our samples, we replace the Si-SiO$_{2}$ interface of a typical
MOSFET with an interface between H terminated Si(111) and a vacuum dielectric in
order to remove the effects of strain and dangling bonds from the
Si(111) surface \cite{eng2007integer}. An encapsulating silicon-on-insulator piece is bonded via
van der Waals forces to the high resistivity H-Si(111) substrate ($<0.5^{\circ}$ miscut) and
forms a gate. The sample studied in the present work is fabricated
in the same way as previous ones, with further optimized cleaning
and annealing steps \cite{mcfarland2009temperaturedependent}. Four-terminal
resistance in a square van der Pauw geometry is defined as $R_{ij,lm}=V_{lm}/I_{ij}$,
with $R_{xx}=R_{12,34}$ and $R_{yy}=R_{24,13}$ oriented along the
{[}1$\bar{1}$0{]} and {[}11$\bar{2}${]} directions respectively
(see inset to Fig.~\ref{fig:bSweep}a). Hall traces, $\rho_{xy}$,
are the averages of $R_{14,23}$ and $R_{23,14}$ to prevent mixing.
The data were obtained at densities between $n_{s}=3.7$ and $\unit[5.7\times10^{11}]{cm^{-2}}$,
adjustable via a gate voltage, in both a $^{3}$He system and $^{3}$He/$^{4}$He
dilution system with base temperatures of $\unit[280]{mK}$ and $\unit[90]{mK}$,
respectively. The density range is limited at high $n_{s}$ due to gate leakage 
and at low $n_{s}$ due to nonlinear contact resistances 
and a small parallel conductance channel, which 
are responsible for non-zero minima at some integer filling factors. Magnetotransport
measurements were performed using standard lock-in techniques with
a typical excitation of $1-\unit[25]{nA}$ at $\unit[5]{Hz}$ in a
magnetic field $B$ up to $\unit[12]{T}$. 

Figure \ref{fig:bSweep}a shows $R_{xx}$, $R_{yy}$ and
$\rho_{xy}$ versus $B$ taken at $T=\unit[90]{mK}$
and density $n_{s}=\unit[3.75\times10^{11}]{cm^{-2}}$. Shubnikov-de
Haas oscillations (SdHO) are visible down to about $B\sim\unit[0.15]{T}$.
The differences in the position of the 
minima at low $B$ between $R_{xx}$ and $R_{yy}$ are not an indication 
of density inhomogeneity; high $B$ minima of $R_{xx}$ and $R_{yy}$ 
are consistent to $0.01\%$ at $\nu=8/5$. While $R_{xx}\approx R_{yy}$ at $B=0$, 
strong anisotropy appears
for $B>0$. Such anisotropy is to be expected in Si(111) when valleys
with different mass tensors have an unequal density of states (DOS)
at $E_{F}$. Here, we simply note that the anisotropy is essentially 
constant above $B=\unit[7]{T}$, and we will focus our analysis 
on these higher magnetic fields. By $B\approx\unit[1.3]{T}$ ($\nu=12$), only the
lowest Landau level is occupied. As $B$ increases, the six-fold valley
degeneracy breaks. Below $\nu=8$ all integer filling factors have
minima, with the even states ($\nu=6,\,4,$ and $2$) being stronger
than the odd ($\nu=3,\,5$ and $7$).

\begin{figure}
	\centering{}\includegraphics[width=3.375in]{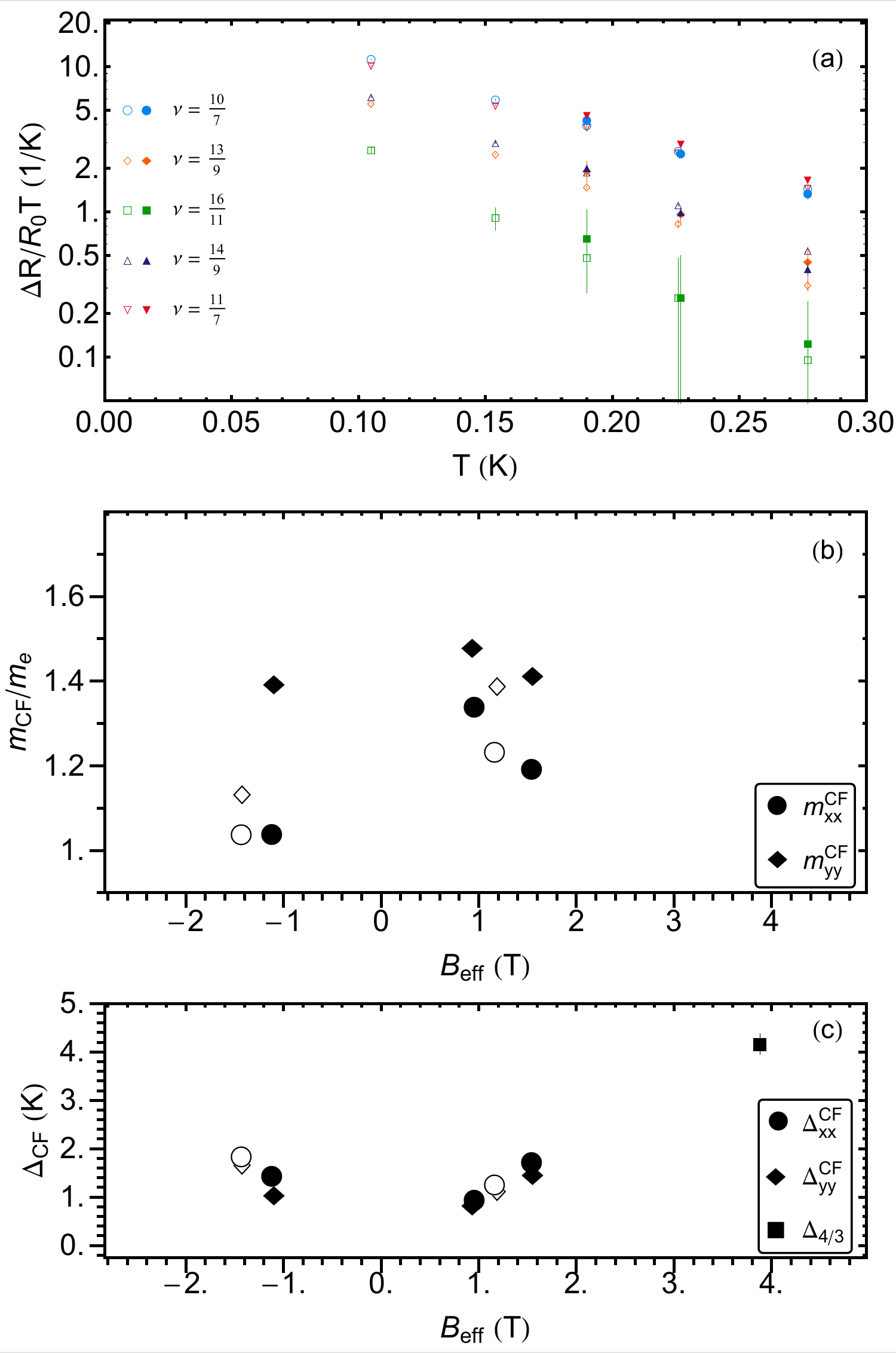}
	\caption{
		(Color) 
		\emph{(a)} Resistance data (with measurement errors) versus temperature
		used in Shubnikov-de Haas analysis. Open (closed) symbols are from $R_{xx}$ ($R_{yy}$).
		Note that only three data points are available for all $R_{yy}$ data.  
		\emph{(b)} Mass of composite fermions assuming that the gaps probed by Shubnikov-de
		Haas (SdH) oscillations have the form $\Delta_{\text{CF}}=\hbar\omega_c^{\text{CF}}$.  
		Open symbols indicate values extracted from maxima. 
		See text for a discussion of the statistical and systematic errors 
		involved in this measurement.
		\emph{(c)} Equivalent CF gap energy calculated from the masses. 
		The gap value at $\nu=4/3$ is calculated 
		from activation energy data and is approximately the same for 
		both $R_{xx}$ and $R_{yy}$. 
		\label{fig:Mass}
	}
\end{figure} 
Figure~\ref{fig:bSweep}b shows the temperature dependence of
$R_{xx}$ and $R_{yy}$ in the FQH regime below $\nu=2$. In this range, we observe
minima at 9 fractions: $\nu=8/5$, 14/9, 20/13, 22/15, 16/11, 10/7, 4/3, (shown) as well
as 6/5 and 14/11 at lower densities (not shown).
Hall plateaus at $\nu=8/5$ and 4/3 are quantized to within 0.5\% of their nominal
values. For weaker fractions, $\nu$ is determined from $B$ evaluated at the 
resistance minimum relative to the value of $B$ at the sharp minimum of 8/5. Using
this technique all fractions deviate less than 0.1\% from their designated values.

The hierarchy of observed states is
exactly that predicted by an SU(2) symmetry in which the two-fold
valley degeneracy of electrons is preserved, leading to the creation
of composite fermions (CF) with two and four attached vortices ($^{2}$CF
and $^{4}$CF, respectively) \cite{park2001thespin}. 
For the hierarchy of fractional states around $\nu=3/2$,
which are the hole-symmetric equivalents to the $\nu=1/2$ states,
the filling factor $\nu^{*}$ of CFs is given by $\nu=2-\nu^{*}/(2p\nu^{*}\pm1)$ \cite{park2001thespin}
where $p=1$ and 2 for $^{2}$CF and $^{4}$CF, respectively.
With an SU(2) symmetry, only $\nu^{*}=2,4,6,\ldots$ are expected,
giving rise to the hierarchy visible in Fig.~\ref{fig:bSweep}b. A
similar, though smaller, hierarchy was observed recently in graphene
with the same conclusion \cite{feldman2012unconventional2}. Finally,
$\nu=6/5$ and 14/11 relate to $^{4}$CFs via $\nu^{*}=4/3$ and 8/5,
respectively, using $2-\nu^{*}/(2\nu^{*}-1)$; the $^{4}$CF hierarchy
simply reflects the FQH states of $^{2}$CFs. We note that we observe
no evidence for a $\nu=5/3$ state, similar to recent experiments
in both graphene \cite{feldman2012unconventional2} and Si/SiGe \cite{lu2012fractional}.
However, the $5/3$ state is visible in nominally doubly-degenerate
AlAs, probably due to local strains 
\cite{padmanabhan2010ferromagnetic,shkolnikov2005observation,abanin2010nematic}.

To estimate the mass of the CFs, we use the theory
of SdHO and apply it to $\nu=3/2$ by transforming to an effective
magnetic field $B_{\text{eff}}=3\left(B-B_{3/2}\right)$ where $B_{3/2}=\unit[10.34]{T}$
(calculated from the density) \cite{du1996composite}. Although we
could use either activation energy measurements or SdHO to find the
gaps, analysis based on SdHO takes into account intrinsic level broadening.
We first compute the amplitude $\Delta R$ of the oscillations using
a linear interpolation of the minima and maxima (see \cite{padmanabhan2008effective}).
We then use $\Delta R\propto R_{0}\exp\left\{ -\pi\Gamma/\Delta\right\} \xi/\sinh\xi$,
where $\xi=2\pi^{2}T/\Delta$ and $R_{0}$ is the (temperature dependent)
resistance at $B_{\text{eff}}=0$, to find the gap $\Delta$ ($\Gamma$ quantifies 
the intrinsic level broadening of Landau levels). 

The extracted SdH data is shown in Fig.~\ref{fig:Mass}a; the vertical bars show the
measurement errors in resistance minima. Note that due to temperature constraints,
there are only three data points available for $R_{yy}$ data -- this means that
a straightforward comparison of masses extracted from $R_{yy}$ and $R_{xx}$ data 
will not have the same precision. Additionally, it is important to note that while
we extract two masses ($m^{\text{CF}}_{xx}$ and $m^{\text{CF}}_{yy}$), we do
not claim that these are aligned to the principal axes of the effective mass
tensor (due to the van der Pauw geometry). Therefore, we cannot make any statement
about the possible anisotropy of composite fermion masses as is argued by
Gokmen et al\cite{gokmen2010transference}.

We expressly assume that $\Delta_{\text{CF}}=\hbar\omega_{c}^{\text{CF}}$,
$\omega_{c}^{\text{CF}}=e\left|B_{\text{eff}}\right|/m_{\text{CF}}$
and extract $m^{\text{CF}}_{xx,yy}$ directly. The results are shown in Fig.~\ref{fig:Mass}b. 
There are two distinct errors that are not shown. First is the statistical error due to the
fitting process. For all cases but one, this is less than 10\%, and for all $m^{\text{CF}}_{xx}$ 
values it is less than 3\%. The second error is the error due to lack of data. While harder
to quantify, it is clear that the $m^{\text{CF}}_{xx}$ values, based on five temperature
values, are more robust than the $m^{\text{CF}}_{yy}$ values. 

By considering the extracted SdH gap energies, however, we note that the 
values for both directions are consistent
with activation energy data; in Fig.~\ref{fig:Mass}c, we
plot the gap energy $\Delta_{\text{CF}}$ as a function of $B_{\text{eff}}$.
The activation energy at $\nu=4/3$, measured from higher $T$ data, is consistent with a constant
value for $m_{\text{CF}}$.  Additionally, we find that all of the
FQH data for $R_{xx}$ collapse onto a single line for $\Delta R\sinh\xi/R_{0}\xi$
vs.~$1/B$ with $\Gamma^{\text{CF}}=\unit[0.83\pm0.04]{K}$, somewhat  
less than twice as large as the value for electrons, $\Gamma=\unit[0.49\pm0.02]{K}$,
determined at low $B$ using the same approach. 

\begin{figure}[t!]
	\centering{}\includegraphics[width=3.375in]{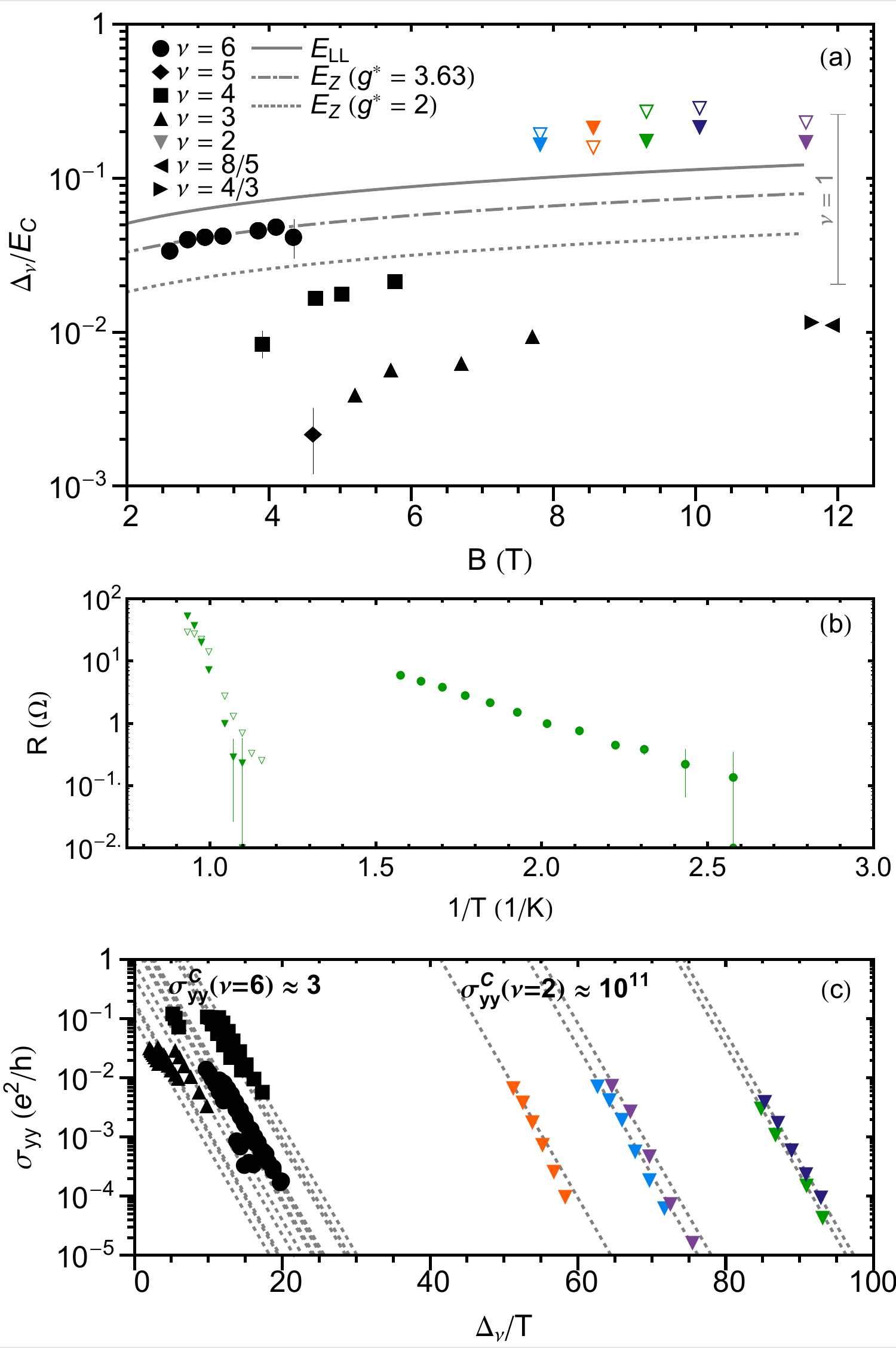}
	\caption{(Color) 
		\emph{(a)} Activation energy measurements as a function of $B$ in
		units of Coulomb energy. Vertical bars show the $\unit[95]{\%}$ confidence
		errors from fits to the linear portions of Arrhenius plots. Closed (open)
		symbols indicate data from $R_{yy}$ ($R_{xx}$) measurements. 
		The $\nu=2$ data shows an enhanced energy
		gap greater than the Landau level spacing ($E_{LL}$) and indicates the scatter
		between $R_{xx}$ and $R_{yy}$ data. 
		\emph{(b)} Arrhenius plot of $R$ for $\nu=2$ and $6$ at 
		$n_{s}=\unit[4.74\times10^{11}]{cm^{-2}}$. While
		the $\nu=6$ dependence is linear over a wide range of temperatures, the
		$\nu=2$ data show a very narrow linear region.
		Both $R_{xx}$ (open symbols) and $R_{yy}$ (closed) show this effect for $\nu=2$. 
		\emph{(c)} Plot of the linear portion of $\sigma_{yy}$ 
		versus $\Delta_{\nu}/T$, showing the difference between the critical
		conductivities $\sigma_{yy}^{C} = \sigma_{yy}\left(1/T\rightarrow0\right)$ 
		at different filling factors and densities. 
		\label{fig:Activation-energy-measurements}
	}
\end{figure}
For comparison to other systems, we define a normalized CF mass 
$m_{\text{nor}}^{\pm}=m_{\text{CF}}\epsilon/\sqrt{B}$
(in units of $\unit[m_{e}]{T^{-1/2}}$, where the $\pm$ is the sign
of $B_{\text{eff}}$) \cite{halperin1993theoryof}. Using $\epsilon=6.25$ 
($\left(\epsilon_{\text{Si}}+\epsilon_{\text{Vac}}\right)/2$,
a value appropriate at a silicon-vacuum interface
for an electron gas with a perpendicular extent $z_\perp$ small compared
to the interaction distance) a weighted average for $R_{xx}$ data gives 
$m_{\text{nor}}^{-}=2.14\pm0.04$ and $m_{\text{nor}}^{+}=2.40\pm0.05$.
In other materials, $m_{\text{nor}}\approx3.2-3.5$ for GaAs at $\nu=1/2$
(\cite{pan2000effective} and references therein) or $m_{\text{nor}}\approx1.7$
at $\nu=3/2$ \cite{du1996composite}, and $m_{\text{nor}}\approx1.1-1.3$
for ZnO at $\nu=1/2$ \cite{maryenko2012temperaturedependent}. 
Unlike our samples, the data for GaAs experiments appears to be symmetric
around $B_{\text{eff}}=\unit[0]{T}$. Furthermore, the ZnO heterostructures
show a linear dependence on the perpendicular magnetic field. Our
results cannot rule out either a divergence or a linear dependence
on $B$. 
There are several material-specific effects that may affect the CF mass, including 
effective-mass anisotropy \cite{yang2012band}, the finite $z_{\perp}$
 of the electron gas, and Landau level mixing
\cite{gee1996composite,melik-alaverdian1995composite}. The modification
of short-range interactions due to the large dielectric-constant,
$\epsilon$, mismatch at the surface is particularly relevant for
our devices. 

We turn now to gaps at integer filling factors; we use the $T$ dependence
of $R_{xx}$ and $R_{yy}$ to calculate the activation energies via
$R\propto\exp\left(-\Delta_{\nu}/2k_{B}T\right)$. By changing $n_{s}$,
we are able to measure the gap $\Delta_{\nu}$ as a function of $B$:
Figure \ref{fig:Activation-energy-measurements}a shows the results
of such an analysis in units of the Coulomb energy $E_{C}=e^{2}/4\pi\epsilon\epsilon_{0}l_{B}$,
where $l_{B}=\sqrt{\hbar/eB}$ and we use $\epsilon=6.25$. Due
to anisotropy and a lack of large resistance range in $R_{xx}$, we
show only $\Delta_{\nu}^{yy}$ for most of the filling factors; for
$\nu=2$ we show both transport directions as an example of scatter
in the gap energy. 

With the assumption that the valley splitting is smaller than the
Zeeman gap, the $\nu=6$ activation energy is interpreted as $\Delta_{6}=E_{Z}=g^{*}\mu_{B}B_{\text{tot}}$,
from which we estimate $g^{*}=3.63\pm0.06$, consistent with previous
measurements \cite{eng2007integer,mcfarland2009temperaturedependent}.
The presence of minima at all filling factors $\nu\leq6$ suggests
a $B$-dependent valley splitting, similar to that observed in SiGe
heterostructures \cite{weitz1996tiltedmagnetic,wilde2005directmeasurements}.
Indeed, the appearance of odd $\nu<6$ shows that high magnetic fields
increasingly break the two-fold degeneracy of opposite $\vec{k}$
valleys. The estimate at $\nu=5$ is based on a ``strength'' ($S$)
of the state defined as the ratio of the resistance minimum to the
average of the adjacent maxima \cite{lu2012fractional}. From the
$T$ dependence, we can estimate a quasi-gap; the state is weaker
than any other filling factor $\nu\leq6$. For $\nu=1$, we can only
estimate an upper and lower bound while noting that the minimum remains
visible at $T=\unit[1]{K}$ (see Fig.~\ref{fig:Activation-energy-measurements}a).
The odd filling factor gaps, which are much greater than $\Gamma$,
indicate that the splitting is likely due to many-body effects. 

For $\nu=2$, the interpretation of the activation energy is more difficult. 
As shown by the solid line in Fig.~\ref{fig:Activation-energy-measurements}a, 
$\Delta_{\nu=2}$ is greater than the Landau level spacing ($E_{LL}$). To show
the qualitative difference between $\nu=2$ and the much better understood $\nu=6$,
we show the temperature dependence of the resistance for the two minima in
Fig.~\ref{fig:Activation-energy-measurements}b at a density of 
$n_{s}=\unit[4.60\times10^{11}]{cm^{-2}}$. The data 
for $\nu=6$ clearly follow an Arrhenius relationship, 
while the resistance change for $\nu=2$ occurs over a very limited temperature
change, and has a much smaller range over which it is linear. 
The transition for $\nu=2$ occurs in a range of $\sim\unit[200]{mK}$ near 
$T\approx\unit[1]{K}$. Another illustration of the peculiarity of $\nu=2$ is
shown in Fig~\ref{fig:Activation-energy-measurements}c, where the thermally
activated portion of $\sigma_{yy}$ as a function of $\Delta_{\nu}/T$ is plotted
for integer filling factors. The critical conductivity
 $\sigma_{yy}^{C}=\sigma_{yy}\left(1/T\rightarrow0\right)$
typically observed in the IQH regime is $\sim2e^{2}/h$, with reductions 
possible due to short-range scattering and screening
\cite{dtextquoterightambrumenil2011modelfor,polyakov1995universal,katayama1994experimental}.
In our data we observe that the criticial conductivity
at $\nu=2$ is 10 to 16 orders of magnitude larger 
than $\sigma_{yy}^{C}$ at higher filling factors.
We therefore consider other interpretations for 
the large activation energy.

At $\nu=2$ one pair of valleys is filled, so added electrons
must occupy a new valley with a different mass tensor. In this regime,
Abanin et al.~\cite{abanin2010nematic} predict a novel nematic
phase with the electron gas broken up into domains of differing
valley polarization. Indeed,
it is characterized by a very narrow $T$ range 
where $R$ is thermally activated, leading to a large extrapolated $R$
for $1/T\rightarrow0$ in an Arrhenius plot of the data. 
Abanin et al.~argue that in an anisotropic valley-degenerate system, the
transport mechanism in the limit of valley-polarized domains is variable-range
hopping due to the low $T$ localization of edge currents along domain
walls separating areas of different valley polarization. Fits of
the temperature dependent resistance to different
models, including $\sigma=\left(\sigma^{C*}/T\right)\exp\left\{ \Delta_{\nu}/2k_{B}T\right\} $,
reproduce the large energy gap and do not reduce the discrepancy in
$\sigma_{yy}^{C}$ \cite{katayama1994experimental}. To the best of
our knowledge, the thermal behavior reported here for $\nu=2$ has
not been seen in any other QHE state. 

In summary, we have shown evidence of electron-electron interactions
in a high mobility Si(111) system. The six-fold valley degeneracy
breaks at high magnetic fields into an apparent SU(2) symmetry reflected
by the fractional quantum Hall state hierarchy. 
While the SU(2) symmetry is not unexpected due to the underlying valley structure of Si(111),
the extended hierarchy reiterates the need to fully understand the valley degeneracy breaking
mechanism in this system. We estimated the
mass of composite fermions near $\nu=3/2$ by assuming that the Shubnikov-de
Haas oscillation gaps can be interpreted as a cyclotron energy. Further
experiments are necessary. First,
hexagonal samples would allow unambiguous measurements of $\rho$
on Si(111) surfaces. Second, higher magnetic fields are necessary
to probe the $\nu<1$ regime at similar densities, which would shed
further light on the valley degeneracy of composite fermions. Finally,
tilted magnetic fields would introduce controlled valley splitting
of inequivalent valleys \cite{eng2007integer,gokmen2008parallel}
and in particular would help shed light on the behavior at $\nu=2$. 
\begin{acknowledgments}
This work was funded by the Laboratory for Physical Sciences. The
authors thank Jainendra Jain for useful discussions. 
\end{acknowledgments}
\bibliographystyle{apsrev4-1}
\bibliography{bibtex}

\end{document}